\documentclass[fleqn,twoside,twocolumn,nofootinbib]{revtex4} % Specifies the document class %,unsortedaddress
\usepackage[sec]{ujp} % \usepackage[cyr]{ujp} for cyrillic, \usepackage[web]{ujp} for web
\begin{document}
\title[Can layered-structure effects be observed]%колонтитул
{Can layered-structure effects be observed,\\ if the Fermi surface is closed?}%
\author{P.V. Gorskyi}%1 автор
\affiliation{Yu. Fed'kovych Chernivtsi National University}%институт
\address{2, Kotsyubyns'kyi Str., Chernivtsi 58012, Ukraine}%адрес
\email{gena_grim@mail.ru}%e-mail
\udk{533.9}  \razd{\secvii}

\setcounter{page}{1296}%
\maketitle

\begin{abstract}
By analyzing the longitudinal conductivity in a quantizing magnetic
field directed perpendicularly to the crystal lattice layers, it has
been demonstrated that the layered-structure effects can be observed not
only in crystals with highly open Fermi surfaces, as was conventionally
believed earlier, but also in crystals with closed ones. The
calculations were carried out in the constant-relaxation-time
approximation. In weak magnetic fields, layered-structure effects
manifest themselves as a phase retardation of Shubnikov--de Haas
oscillations and a certain increase of the relative contribution
made by the latter. In the range of high magnetic fields, there
exists an optimal interval, in which the layered-structure effects
reveal themselves in the form of a sharp non-monotonous dependence of
conductivity on the magnetic field. In addition, it has been shown that
the layered-structure effects result in a decrease of the proportionality
factor between the magnetoresistance and the magnetic induction in
the longitudinal Kapitsa effect. The longitudinal conductivity of
layered crystals in ultra-quantum magnetic fields has also been
analyzed. It is shown that the following dependences of the
magnetoresistance on the magnetic field can be obtained, depending
on the model used for the filling of the single Landau subband and
on whether the longitudinal conductivity is considered to be of
either the drift or diffusion type: $\rho_{zz}\propto TB^{2}$,\
$\rho_{zz}\propto B^{3}$, and $\rho_{zz}\propto B^{4}$.
\end{abstract}

\section{Introduction}

The model of band spectrum for charge carriers in layered crystals
was proposed by R.F.~Fivaz as early as in 1967 \cite{1}. Since
then, a lot of works devoted to theoretical and experimental
researches of various physical characteristics of those crystals
have been published. However, it has been considered till now that
the layered-structure effects in electronic processes that take
place in those crystals are pronounced only if the corresponding
Fermi surface (FS) is open, i.e. the FS occupies the whole first
Brillouin zone and, when being periodically continued, forms a
connected surface \cite{2,3,4,5}. On the contrast, crystals with
closed FSes---i.e. FSes which occupy only some part of the first
Brillouin zone and turn out unconnected at the periodic
continuation--are not classified as layered ones by the majority
of researchers, even if the motion of charge carriers along the
superlattice axis (i.e. perpendicularly to the layers) is
described by the strong-coupling law rather than the effective
mass one. At the same time, in the works published earlier
\cite{6,7,8,9}, the analysis of the diamagnetic susceptibility of
an electron gas showed that the layered-structure effects can
manifest themselves, if the FS is closed too, provided that the
level of the filling of a miniband is high. This work aimed at
studying the influence of layered-structure effects on the
longitudinal conductivity of layered crystals with closed FSes in
a strong quantizing magnetic field. The problem concerned is
examined in the approximation of Ohm's law applicability to the
case where a strong quantizing magnetic field and an electric
field are parallel to each other and perpendicular to the layer
planes.

\section{Calculation of Longitudinal Electroconductivity in the Layered
Crystal and Discussion of the Results Obtained}

To calculate the longitudinal conductivity of a layered crystal in a
quantizing magnetic field directed perpendicularly to the planes of layers, the
following law for the charge carrier dispersion was used:
%1
\begin{equation}
\varepsilon\left(  n,x\right)  =\mu^{\ast}B\left(  2n+1\right)  +W\left(
x\right).\label{GrindEQ__1_}%
\end{equation}
Here, $\mu^{\ast}=\mu_{\rm B}{m_{0}/m^{\ast}}$, $\mu_{\rm B}$ is the
Bohr magneton, $m_{0}$ the free electron mass, $m^{\ast}$ the
effective mass of electron in the layer plane, $B$ the magnetic
field induction, $n$ the number of a Landau level, $W(x)$ the
dispersion law for charge carriers along the superlattice axis,
$x=ak_{z}$, $k_{z}$ is the quasimomentum component along the
superlattice axis, and $a$ the distance between translationally
equivalent layers.

To study the influence of layered-structure effects on the longitudinal
electroconductivity, the latter was calculated in two cases, namely,
in the case of the strong-coupling approximation,
%2
\begin{equation}
W\left(  x\right)  =\Delta\left(  1-\cos x\right),\label{GrindEQ__2_}%
\end{equation}
where $\Delta$ is the miniband halfwidth which governs the electron motion
between layers; and in the case where the right-hand side in formula
(\ref{GrindEQ__2_}) is expanded into a series in $x$ up to the quadratic term,
which corresponds to the effective mass approximation. In both cases, the
dependence of the chemical potential in the electron gas on the magnetic field
induction was taken into account. To simplify calculations, the relaxation
time of charge carriers was assumed to be constant. With the same purpose in
view, the influence of the Dingle factor on longitudinal conductivity oscillations
(Shubnikov--de Haas oscillations) was not considered in detail, although it
can be substantial at the electron scattering by impurities and defects in the
crystal lattice \cite{5,10}.

In work \cite{11}, a detailed derivation was given for the equation of
chemical potential in the electron gas in a quantizing magnetic field, as well
as the formulas for components of the longitudinal electroconductivity, in the
case of an arbitrary function $W(x)$ and provided that the relaxation time
$\tau$ of charge carriers is either a constant or a function of the
longitudinal quasimomentum only, i.e. $\tau=\tau\left(  x\right)  $. Then, if
Shubnikov--de Haas oscillations are fierce, the longitudinal crystal
electroconductivity can be determined as follows:
%3
\begin{equation}
\sigma_{zz}\left(  B\right)  =\sigma_{0}+\sigma_{os}\left(  B\right)
.\label{GrindEQ__3_}%
\end{equation}
The individual components in Eq.~((\ref{GrindEQ__3_}) are given by the
formulas \cite{11}
%4
\begin{equation}
\sigma_{0}=\frac{16\pi^{2}e^{2}m^{\ast}a}{h^{4}}\int\limits_{W\left(
x\right) \leq\zeta}\tau\left(  x\right)  \left\vert W^{\prime}\left(
x\right)
\right\vert ^{2}dx,\label{GrindEQ__4_}%
\end{equation}
%
%5
\[
\sigma_{os} \left(  B\right)  =\frac{32\pi^{2} e^{2} m^{*} a}{h^{4} }
\sum_{l=1}^{\infty}\left(  -1\right)  ^{l} f_{l}^{\sigma} \times
\]
\begin{equation}
\label{GrindEQ__5_}\times\int\limits_{W\left(  x\right)
\le\zeta}\tau\left( x\right)  \left|  W^{\prime}\left(  x\right)
\right|  ^{2} \cos\left(  \pi l\frac{\zeta-W\left(  x\right)
}{\mu^{*} B} \right) dx,
\end{equation}
%
%6
\begin{equation}
f_{l}^{\sigma}=\frac{{\pi^{2}%
lkT\mathord{\left/{\vphantom{\pi ^{2} lkT \mu ^{*} B}}\right.\kern-\nulldelimiterspace}\mu
^{\ast}B}}{sh\left(  {\pi^{2}%
lkT\mathord{\left/{\vphantom{\pi ^{2} lkT \mu ^{*} B}}\right.\kern-\nulldelimiterspace}\mu
^{\ast}B}\right)}.\label{GrindEQ__6_}%
\end{equation}
The integration in formulas (\ref{GrindEQ__4_}) and
(\ref{GrindEQ__5_}) is carried out over the positive $x$-values
only. The quantity $\zeta$ is the chemical potential of an electron
gas, the other notations either are standard or were explained
above. The equation that determines the chemical potential of an
electron gas in a quantizing magnetic field, provided that the
Shubnikov--de Haas effect is pronounced, looks like
%7
\[
n_{0}=\frac{4m^{\ast}}{ah^{2}}\int\limits_{W\left(  x\right)
\leq\zeta}\left[ \zeta-W\left(  x\right)  \right]  dx+\frac{8\pi
m^{\ast}kT}{ah^{2}}\times
\]
\begin{equation}
\times\sum_{l=1}^{\infty}\frac{\left(  -1\right)  ^{l}}{{\rm
sh}\left( {\pi
^{2}%
lkT\mathord{\left/{\vphantom{\pi ^{2} lkT \mu ^{*}
B}}\right.\kern-\nulldelimiterspace}\mu ^{\ast}B}\right)
}\int\limits_{W\left(  x\right)  \leq\zeta}\sin\left(  \pi
l\frac{\zeta-W\left(  x\right)  }{\mu^{\ast}B}\right),  \label{GrindEQ__7_}%
\end{equation}
where $n_{0}$ is the concentration of charge carriers in the crystal bulk.
Under conditions that the relaxation time is constant, $\tau\left(  x\right)
=\tau_{0}$, dispersion law (\ref{GrindEQ__2_}) is fulfilled, and the FS is
closed (for such an FS, $\zeta<2\Delta$), formulas (\ref{GrindEQ__4_}) and
(\ref{GrindEQ__5_}) read, respectively,
%8
\begin{equation}
\sigma_{0}=\frac{8\pi^{2}e^{2}m^{\ast}a\tau_{0}\Delta^{2}}{h^{4}}\left(
C_{0}-C_{2}\right),\label{GrindEQ__8_}%
\end{equation}%
%9
\[
\sigma_{os}\left(  B\right)  =\frac{16\pi^{2}e^{2}m^{\ast}a\tau_{0}\Delta^{2}%
}{h^{4}}\times
\]
\[
\times\sum_{l=1}^{\infty}\left(  -1\right)  ^{l} f_{l}^{\sigma} \left\{
\cos\left(  \pi l\frac{\zeta-\Delta}{\mu^{*} B} \right)  \left[  \left(  C_{0}
-C_{2} \right)  J_{0} \left(  \frac{\pi l\Delta}{\mu^{*} B} \right)  +\right.
\right.
\]
\[
+\sum_{r=1}^{\infty}\left(  -1\right)  ^{r} \left(  2C_{2r} -C_{2r+2}
-C_{2r-2} \right)  \times
\]
\[
\times\left.  J_{2r} \left(  \frac{\pi l\Delta}{\mu^{*} B} \right)
\right] -\sin\left(  \pi l\frac{\zeta-\Delta}{\mu^{*} B} \right)
\sum_{r=0}^{\infty} (-1)^r\left(  2C_{2r+1} -\right.
\]
\begin{equation}
-C_{2r+3}-\left.  \left.  C_{\left\vert 2r-1\right\vert }\right)
J_{2r+1}\left(  \frac{\pi l\Delta}{\mu^{\ast}B}\right)  \right\}.
\label{GrindEQ__9_}%
\end{equation}
In these formulas,
%10
\begin{equation}
C_{0}=\arccos\left(  1-\gamma\right), \label{GrindEQ__10_}%
\end{equation}
%
%11
\begin{equation}
C_{m}=\frac{\sin mC_{0}}{m}\quad\mathrm{for} \quad m\neq0, \label{GrindEQ__11_}%
\end{equation}%
%12
\begin{equation}
\gamma={\zeta
\mathord{\left/{\vphantom{\zeta \Delta }}\right.\kern-\nulldelimiterspace}\Delta
}, \label{GrindEQ__12_}%
\end{equation}
and $J_{n}(z)$ are the Bessel functions of the first kind of the real argument
$z$.

In the case of a closed FS and the dispersion law (\ref{GrindEQ__2_}),
Eq.~(\ref{GrindEQ__7_}) reads
%13
\[
n_{0}=\frac{4m^{\ast}\Delta}{ah^{2}}\left[  \left(  \gamma-1\right)
C_{0}+\sqrt{2\gamma-\gamma^{2}}\right]  +
\]
\[
+\frac{8m^{*} \pi kT}{ah^{2} } \sum_{l=1}^{\infty}\frac{\left(  -1\right)
^{l} }{sh\left(  {\pi^{2} lkT \mathord{\left/{\vphantom{\pi
^{2} lkT \mu ^{*} B}}\right.\kern-\nulldelimiterspace} \mu^{*} B} \right)  }
\left\{  \sin\left(  \pi l\frac{\zeta-\Delta}{\mu^{*} B} \right)
\times\right.
\]
\[
\times\left[  C_{0} J_{0} \left(  \frac{\pi l\Delta}{\mu^{*} B} \right)
+2\sum_{r=1}^{\infty}\left(  -1\right)  ^{r} C_{2r} J_{2r} \left(  \frac{\pi
l\Delta}{\mu^{*} B} \right)  \right]  +
\]
\begin{equation}
\left.  +\mathrm{\;2cos}\left(  \pi l\frac{\zeta-\Delta}{\mu^{\ast}B}\right)
\sum_{r=0}^{\infty}\left(  -1\right)  ^{r}C_{2r+1}J_{2r+1}\left(  \frac{\pi
l\Delta}{\mu^{\ast}B}\right)  \right\}. \label{GrindEQ__13_}%
\end{equation}
In the case of an open FS, i.e. if $\gamma\geq2$, one has to put $C_{0}=\pi$ in
formulas (\ref{GrindEQ__8_})--(\ref{GrindEQ__13_}) and, additionally, put the
radical in formula (\ref{GrindEQ__13_}) to be equal zero.

By passing to the effective mass approximation in formulas (\ref{GrindEQ__8_}), (\ref{GrindEQ__9_}), and
Eq.~(\ref{GrindEQ__13_}), we obtain,
respectively,
%14
\begin{equation}
\sigma_{0}=\frac{16\pi^{2}e^{2}m^{\ast}a\tau_{0}\Delta^{2}}{3h^{4}}\left(
\frac{2\zeta}{\Delta}\right)
^{{3\mathord{\left/{\vphantom{3 2}}\right.\kern-\nulldelimiterspace}2}},
\label{GrindEQ__14_}%
\end{equation}%
%15
\[
\sigma_{os}\left(  B\right)  =\frac{32\pi^{{1\mathord{\left/{\vphantom{1
2}}\right.\kern-\nulldelimiterspace}2}}e^{2}m^{\ast}a\tau_{0}\Delta
^{{1\mathord{\left/{\vphantom{1 2}}\right.\kern-\nulldelimiterspace}2}}\left(
\mu^{\ast}B\right)  ^{{3\mathord{\left/{\vphantom{3
2}}\right.\kern-\nulldelimiterspace}2}}}{h^{4}}\times
\]
\[
\times\sum_{l=1}^{\infty}\frac{\left(  -1\right)  ^{l} f_{l}^{\sigma} }{l^{{3
\mathord{\left/{\vphantom{3 2}}\right.\kern-\nulldelimiterspace} 2} } }
\left[  \sin\left(  \frac{\pi l\zeta}{\mu^{*} B} \right)  C\left(  \sqrt
{\frac{2l\zeta}{\mu^{*} B} } \right)  -\right.
\]
\begin{equation}
\label{GrindEQ__15_}\left.  -\cos\left(  \pi l\frac{\zeta-\Delta}{\mu^{*} B}
\right)  S\left(  \sqrt{\frac{2l\zeta}{\mu^{*} B} } \right)  \right],
\end{equation}
%
%16
\[
n_{0} =\frac{8m^{*} \zeta}{3ah^{2} } \sqrt{\frac{2\zeta}{\Delta} } +\frac{8\pi
m^{*} kT}{ah^{2} } \sqrt{\frac{\mu^{*} B}{\Delta} } \times
\]
\[
\times\sum_{l=1}^{\infty}\frac{\left(  -1\right)  ^{l} }{l^{{1
\mathord{\left/{\vphantom{1 2}}\right.\kern-\nulldelimiterspace} 2}
} {\rm sh}\left(  {\pi^{2} lkT \mathord{\left/{\vphantom{\pi ^{2}
lkT \mu ^{*} B}}\right.\kern-\nulldelimiterspace} \mu^{*} B} \right)
} \left[  \sin\left( \frac{\pi l\zeta}{\mu^{*} B} \right)  C\left(
\sqrt{\frac{2l\zeta}{\mu^{*} B} } \right)  \right.  +
\]
\begin{equation}
\left.  +\cos\left(  \frac{\pi l\zeta}{\mu^{\ast}B}\right)  S\left(
\sqrt{\frac{2l\zeta}{\mu^{\ast}B}}\right)  \right]. \label{GrindEQ__16_}%
\end{equation}
In formulas (\ref{GrindEQ__15_}) and (\ref{GrindEQ__16_}), the functions
$C(z)$ and $S(z)$ are the Fresnel cosine and sine integrals, respectively. The
charge carrier concentration was assumed constant in both examined cases. It
was determined by the formula
%17
\begin{equation}
n_{0}=\frac{4m^{\ast}\Delta}{ah^{2}}\left[  \left(  \gamma_{0}-1\right)
\arccos\left(  1-\gamma_{0}\right)  +\sqrt{2\gamma_{0}-\gamma_{0}^{2}}\right],
\label{GrindEQ__17_}%
\end{equation}
where $\gamma_{0}={\zeta_{0}/\Delta}$, and $\zeta_{0}$ is the Fermi energy of
the electron gas in the crystal, provided the dispersion law (\ref{GrindEQ__2_}),
the zero absolute temperature, and the absence of a magnetic field.

Before passing to a more detailed analysis of the influence of
layered-structure effects on the longitudinal electroconductivity of the
crystal, we note that, in the quasiclassical approximation, when the conditions
${\Delta/(\mu^{\ast}B)\gg1}$ and ${\zeta/(\mu^{\ast}B)\gg1}$ are satisfied,
formulas (\ref{GrindEQ__9_}) and (\ref{GrindEQ__15_}) for the oscillating part
of the longitudinal electroconductivity give the same result,
%18
\[
\sigma _{os} \left(B\right)=\frac{16\sqrt{2} \pi ^{{1
\mathord{\left/{\vphantom{1 2}}\right.\kern-\nulldelimiterspace} 2}
}
 e^{2} m^{*} a\tau _{0} \Delta ^{{1 \mathord{\left/{\vphantom{1 2}}\right.\kern-\nulldelimiterspace} 2} } \left(\mu ^{*} B\right)^{{3 \mathord{\left/{\vphantom{3 2}}\right.\kern-\nulldelimiterspace} 2} } }{h^{4} }
 \times\]
 \begin{equation} \label{GrindEQ__18_}
 \times \sum _{l=1}^{\infty }\left(-1\right)^{l} l^{-{3 \mathord{\left/{\vphantom{3 2}}\right.\kern-\nulldelimiterspace} 2} } \sin \left(\frac{\pi l\zeta }{\mu ^{*} B} -\frac{\pi }{4} \right) .
\end{equation}
It is not of surprise, because both the specific dispersion law for
charge carriers and the FS finiteness -- both in the magnetic field
direction -- are insignificant in the quasiclassical approximation.
However, in formulas (\ref{GrindEQ__9_}) and (\ref{GrindEQ__15_}),
those factors were taken into account explicitly. Therefore, in the
framework of this approximation, it is impossible to distinguish the
influence of layered-structure effects on the longitudinal
electroconductivity in crystals with a closed FS, provided that the
Fermi energies are identical.

%Fig. 1
\begin{figure}% figure* for wide figure, [h] [!] to change the placement
\includegraphics[width=6.8cm]{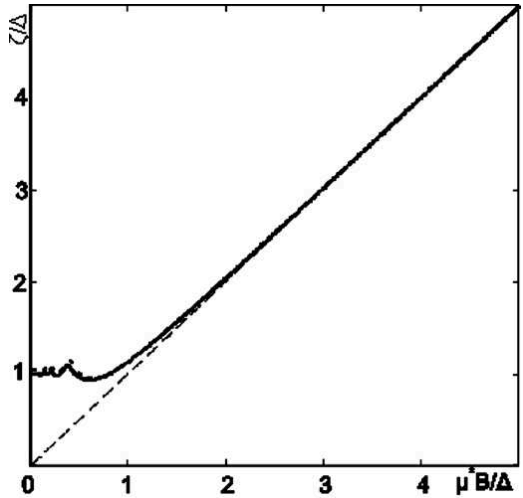}
\vskip-3mm\caption{Field dependences of the chemical potential at
$\gamma_{0}=1$ and ${kT/\Delta=0.03}$ for a layered crystal (solid
curve) and in the effective mass approximation (dashed curve). The
thin dashed line corresponds to the limiting case
$\zeta=\mu^{\ast}B$  }
\end{figure}

However, if the concentration of charge carriers is constant, the
layered-structure effects will affect electroconductivity oscillations even in
the quasiclassical approximation, because the Fermi energy in a weak magnetic
field is a little higher in the effective mass approximation than that in the case
of the dispersion law (\ref{GrindEQ__2_}). It is so because, if a finite width
of the miniband is taken into account explicitly, the density of states turns out
higher than that in the effective mass approximation. The field dependences of
the chemical potential in the electron gas are depicted in Fig.~1 in the range of
magnetic fields $0\leq{\mu^{\ast}B/\Delta\leq5}$, at ${kT/\Delta=0.03}$ and
${\zeta_{0}/\Delta=1}${,} and for the cases of a real layered crystal with
the dispersion law (\ref{GrindEQ__2_}) (solid curve) and in the effective mass
approximation (dashed curve). The dependences were obtained without taking
the Dingle factor into consideration. The difference of the latter from 1
can be neglected, when the condition that the scattering-induced broadening of
energy levels is small in comparison with the distance between Landau levels
is well satisfied within the whole examined range of magnetic fields. This
condition will be discussed below in more details, when estimating the
longitudinal conductivity numerically.

As the magnetic field grows, both curves approach each other, because, in the
case of closed FS, Eq.~(\ref{GrindEQ__13_}) has the following asymptotic
solution in strong quantizing magnetic fields \cite{11}:
%19
\begin{equation}
\zeta\left(  B\right)  =\mu^{\ast}B+\Delta\left[  1-\cos\left(  \frac{f\left(
\gamma_{0}\right)  \Delta}{2\mu^{\ast}B}\right)  \right],
\label{GrindEQ__19_}%
\end{equation}
where
%20
\begin{equation}
f\left(  \gamma_{0}\right)  =\left(  \gamma_{0}-1\right)  \arccos\left(
1-\gamma_{0}\right)  +\sqrt{2\gamma_{0}-\gamma_{0}^{2}}. \label{GrindEQ__20_}%
\end{equation}
Whence, it is evident that the single filled Landau subband becomes narrower
in the ultraquantum limit, and this narrowing has to be taken into account in calculations of
the longitudinal conductivity.

%Fig. 2
\begin{figure}% figure* for wide figure, [h] [!] to change the placement
\includegraphics[width=7cm]{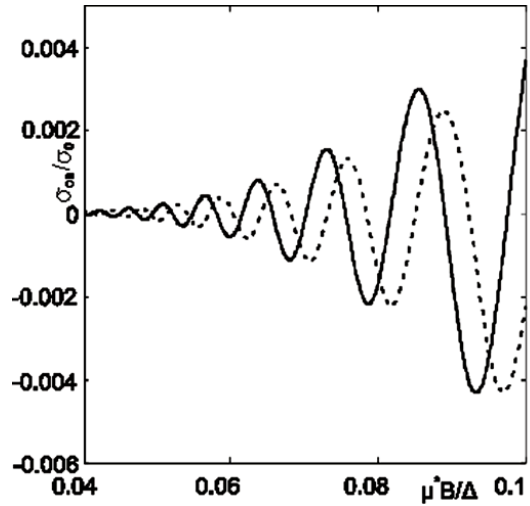}
\vskip-3mm\caption{Field dependences of the oscillating part of the
longitudinal conductivity at $\gamma_{0}=1$ and ${kT/\Delta=0.03}$
for the layered crystal (solid curve) and in the effective mass
approximation (dashed curve)  }
\end{figure}

In the effective mass approximation, formula (\ref{GrindEQ__19_}) reads
%21
\begin{equation}
\zeta\left(  B\right)  =\mu^{\ast}B+\frac{\Delta^{3}f^{2}\left(  \gamma
_{0}\right)  }{8\left(  \mu^{\ast}B\right)  ^{2}}.\label{GrindEQ__21_}%
\end{equation}
Note that all the formulas given above for the longitudinal conductivity
were also obtained in the case where the interaction-induced
broadening of energy levels is small in comparison with the distance
between Landau levels. Only in this case, the broadening of energy
levels can be associated directly with the relaxation time, and the
scattering-induced shift of energy levels can be neglected. Then
the approach based on the Boltzmann equation turns out completely
equivalent to that based on the Kubo formalism. However, in the
first part of work \cite{5}, the opposite case was considered,
namely, when the scattering-induced broadening of energy levels is
large in comparison with the distance between Landau levels.
However, even in this case, the formula obtained in work \cite{4}
for the invariable component of the longitudinal conductivity
coincides--to within an exponentially small (at low
temperatures) correction--with formula (8) derived for the case of
the open FS. The field dependences of the oscillating part in the longitudinal
electroconductivity
are shown in Fig.~2 in the magnetic field range $0.04\leq{\mu^{\ast}%
B/\Delta\leq0.1}$, for ${kT/\Delta=0.03}$ and ${\zeta_{0}/\Delta=1}$, and for
the cases of a real layered crystal with the dispersion law (\ref{GrindEQ__2_})
(solid curve) and in the effective mass approximation (dashed curve). The
figure demonstrates that, owing to some difference between oscillation
frequencies which was mentioned above, oscillations of the longitudinal
electroconductivity in a layered crystal with the dispersion law
(\ref{GrindEQ__2_}) acquire, as the magnetic field induction grows, some phase
retardation with respect to the same oscillations calculated in the effective
mass approximation. Moreover, the relative contribution of oscillations in the
case of the dispersion law (\ref{GrindEQ__2_}) is larger than that in the
effective mass approximation. This occurs because, if layered-structure effects
are considered, the slope of the FS cross-section as a function of the
longitudinal quasimomentum is less than that in the effective mass approximation, and,
in addition, the longitudinal velocity of charge carriers is less in the former
case. The analysis testifies that, if the Fermi energy is assumed
constant, the considered oscillations coincide by their phase and frequency.
However, the relative contribution of the oscillating part is somewhat larger,
if the layered-structure effects are taken into account. In both cases, this
contribution achieves 0.4\% in the range of magnetic fields $0.04\leq
{\mu^{\ast}B/\Delta\leq0.1}$.

%Fig. 3
\begin{figure}% figure* for wide figure, [h] [!] to change the placement
\includegraphics[width=7cm]{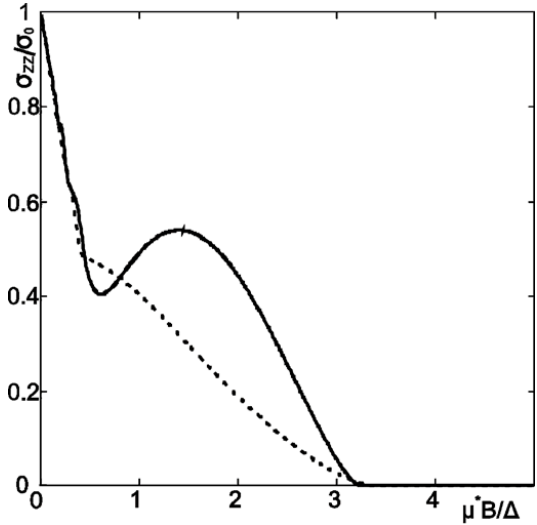}
\vskip-3mm\caption{Field dependences of the total longitudinal
conductivity at $\gamma_{0}=1$ and ${kT/\Delta=0.03}$ for a layered
crystal (solid curve) and in the effective mass approximation
(dashed curve) }
\end{figure}

Now let us calculate the total longitudinal conductivity in a magnetic
field. As was already mentioned in work \cite{11}, one has to
consider the narrowing of the single filled Landau subband.
Therefore, the substitution $\gamma={\left(  \zeta-\mu^{\ast}B\right)
/\Delta}$ has to be done in formulas (\ref{GrindEQ__8_}) and
(\ref{GrindEQ__9_}) with regard for Eqs.~(\ref{GrindEQ__10_})--(\ref{GrindEQ__12_}).
At the same time, formulas (\ref{GrindEQ__14_}) and
(\ref{GrindEQ__15_}) derived in the effective mass approximation look like
%22
\begin{equation}
\sigma_{0}=\frac{32\sqrt{2}\pi^{2}e^{2}m^{\ast}a\tau_{0}\Delta
^{1/2}%
}{3h^{4}}\left(  \zeta-\mu^{\ast}B\right) ^{3/2},
\label{GrindEQ__22_}%
\end{equation}
%
%23
\[
\sigma_{os} \left(  B\right)  =\frac{32\pi^{{1 \mathord{\left/{\vphantom{1
2}}\right.\kern-\nulldelimiterspace} 2} } e^{2} m^{*} a\tau_{0} \Delta^{{1
\mathord{\left/{\vphantom{1 2}}\right.\kern-\nulldelimiterspace} 2} } \left(
\mu^{*} B\right)  ^{{3 \mathord{\left/{\vphantom{3
2}}\right.\kern-\nulldelimiterspace} 2} } }{h^{4} } \times
\]
\[
\times\sum_{l=1}^{\infty}\frac{\left(  -1\right)  ^{l} f_{l}^{\sigma} }{l^{{3
\mathord{\left/{\vphantom{3 2}}\right.\kern-\nulldelimiterspace} 2} } }
\left\{  \sin\left(  \frac{\pi l\zeta}{\mu^{*} B} \right)  C \left[
\sqrt{\frac{2l\left(  \zeta-\mu^{*} B\right)  }{\mu^{*} B} } \right]
-\right.
\]
\begin{equation}
\left.  -\cos\left(  \frac{\pi l\zeta}{\mu^{\ast}B}\right)  S\left[
\sqrt{\frac{2l\left(  \zeta-\mu^{\ast}B\right)  }{\mu^{\ast}B}}\right]
\right\}  . \label{GrindEQ__23_}%
\end{equation}

The field dependences of the total longitudinal electroconductivity in the range
of magnetic fields $0\leq{\mu^{\ast}B/\Delta\leq5}$ and at ${kT/\Delta=0.03}$
and ${\zeta_{0}/\Delta=1}$ are exhibited in Fig.~3 for the cases of a real
layered crystal with the dispersion law (\ref{GrindEQ__2_}) (solid curve) and
in the effective mass approximation (dashed curve). The figure
demonstrates that, in a weak magnetic field, the total conductivity is
slightly different in both considered cases, although the contribution of
the oscillating term is more pronounced if the layered-structure effects are taken
into account. In a very strong magnetic field, they again weakly differ from
each other, because formulas (\ref{GrindEQ__19_}) and (\ref{GrindEQ__21_})
give practically the same result here. However, there exists a certain
\textquotedblleft optimal\textquotedblright\ range of magnetic fields
$0.5\leq{\mu^{\ast}B/\Delta\leq3}$, where the layered-structure effects are
the most pronounced owing to both the slower dependence of the FS
cross-section area on the longitudinal quasimomentum and the smaller
longitudinal velocity of charge carriers. Hence, one can see that
the layered-structure effects can be highly expressed in crystals with a closed FS
as well.

Consider now, for example, the Kapitsa longitudinal effect \cite{12} which
consists in the linear dependence of the longitudinal magnetoresistance on the
magnetic field induction. Expanding Eq.~(\ref{GrindEQ__8_}) in a power
series in the small parameter ${\mu^{\ast}B/\Delta}$ and confining the
expansion to linear terms, we obtain the following formula for the relative
magnetoresistance:
%24
\begin{equation}
\frac{\Delta\rho\left(  B\right)  }{\rho\left(  0\right)  }=\frac{2\sin C_{0}%
}{C_{0}-C_{2}}\frac{\mu^{\ast}B}{\Delta}.\label{GrindEQ__24_}%
\end{equation}
For instance, at $\gamma=1$ and ${\mu^{\ast}B/\Delta=0.1}${, }taking into
account that $C_{0}={\pi/2}${ and }${C_{2}=0}$, we obtain a value of 12.7\%
for the relative magnetoresistance. At the same time, in the effective mass
approximation, by expanding expression (\ref{GrindEQ__22_}) in a Taylor
series, we obtain the formula
%25
\begin{equation}
\frac{\Delta\rho\left(  B\right)  }{\rho\left(  0\right)  }=1.5\sqrt{\gamma
}\frac{\mu^{\ast}B}{\Delta}.\label{GrindEQ__25_}%
\end{equation}
Under the conditions given above, we obtain a value of 15\% for the relative
magnetoresistance, i.e. a little larger than in the case where the layered-structure
effects were taken into account. From the presented results, it follows that,
in the range of magnetic fields $0\leq{\mu^{\ast}B/\Delta\leq0.1}$, the
contribution of the Kapitsa effect to the longitudinal magnetoresistance of a
crystal is much larger than that of the Shubnikov--de Haas one.

Note that it is impossible to explain the Kapitsa longitudinal effect within the
conventional approaches. Those approaches can explain only the
transverse Kapitsa effect, and its physical reason is understood as if the
role of mean free paths of charge carriers in strong magnetic fields is
gradually transferred to the radii of their cyclotron orbits \cite{13,14}.

Now let us estimate the longitudinal electroconductivity of the crystal in the
absence of a magnetic field, because it is a reference point for the
characteristics depicted in Figs.~2 and 3. Consider the region of the charge
carrier scattering at charged impurities. Since the relaxation time is assumed
to be constant, it can be determined by the formula
%26
\begin{equation}
\tau_{0}=\frac{2\pi lm_{es}^{\ast}}{hk_{0}}.\label{GrindEQ__26_}%
\end{equation}
In this expression, $l$ is the mean free path of charge carriers; $k_{0}$ the
equivalent radius of a Fermi sphere that simulates the real FS in the layered
crystal to consider the charge carrier scattering as isotropic; and $m_{es}^{\ast
}$ the effective mass of charge carriers on that sphere. The formulas for
those quantities were obtained using the conditions that the Fermi energy and
the concentration of charge carriers are the same as in a real crystal. At
$\gamma=1$, these are
%27
\begin{equation}
k_{0}=\sqrt[3]{\frac{12\pi^{2}m^{\ast}\Delta}{ah^{2}}},\label{GrindEQ__27_}%
\end{equation}
%
%28
\begin{equation}
m_{es}^{\ast}=\frac{h^{2}k_{0}^{2}}{8\pi^{2}\Delta}. \label{GrindEQ__28_}%
\end{equation}
Substituting Eqs.~(\ref{GrindEQ__27_}) and (\ref{GrindEQ__28_}) into
Eq.~(\ref{GrindEQ__8_}), taking into account that $\gamma=1$, and introducing
the notation $N={l/a}$, we obtain the following final formula for the
longitudinal conductivity of a layered crystal in the absence of a magnetic
field:
%29
\begin{equation}
\sigma_{0}=\frac{2\pi e^{2}m^{\ast}a^{2}\Delta}{h^{3}}\sqrt[3]{\frac{12\pi
^{2}m^{\ast}\Delta}{ah^{2}}}N. \label{GrindEQ__29_}%
\end{equation}
The corresponding analysis showed that, for the parameters
$a=1$\textrm{~nm}, $m^{\ast}=0.01m_{0}$, and $\Delta=0.01~\mathrm{eV}$, the
Dingle factor does not substantially influence the chemical potential
oscillations and the oscillating part of longitudinal conductivity, if
$N\geq1000$. Hence, in accordance with formula (\ref{GrindEQ__29_}), the
minimum value of longitudinal conductivity in the layered crystal under given
conditions can amount to $1.06\times10^{3}~\Omega^{-1}\mathrm{m}^{-1}$. In the
effective mass approximation, for the same relaxation time and charge carrier
concentration, this conductivity is about 20\% higher and amounts to
$1.272\times10^{3}~\Omega^{-1}\mathrm{m}^{-1}$. This result has a simple
physical interpretation: any restriction imposed on the free motion of charge
carriers in the direction perpendicular to the layers reduces the conductivity
of the crystal in this direction.

At last, let us obtain an explicit asymptotic expression for the
electroconductivity in a layered crystal in strong ultraquantum magnetic
fields. For this purpose, let us substitute the quantities $\tau_{0}$ from
Eq.~(\ref{GrindEQ__26_}) and $\zeta\left(  B\right)  $ from
Eq.~(\ref{GrindEQ__21_}) into formula (\ref{GrindEQ__5_}) immediately. Then,
the multiplier $\left(  -1\right)  ^{l}$ under the sum over $l$ is
compensated, the cosine can be substituted by 1, and the integration over
$x$ can be executed within the limits from 0 to $\frac{f\left(  \gamma
_{0}\right)  \Delta}{2\mu^{\ast}B}$. When integrating, the square of
the longitudinal velocity of charge carriers in the integrand has to be replaced
by its principal term of the expansion in $x$, i.e. the quadratic one. In addition, the numerical
analysis demonstrates that, at ${\mu^{\ast}B/kT\gg1}$,
%30
\begin{equation}
\sum_{l=1}^{\infty}f_{l}^{\sigma}=\frac{2.467\mu^{\ast}B}{\pi^{2}%
kT}.\label{GrindEQ__30_}%
\end{equation}
Executing all the quoted transformations and combining all numerical
multipliers into a single one, we obtain the following final asymptotic
expression for the longitudinal conductivity:
%31
\begin{equation}
\sigma_{zz}\left(  B\right)  =1.285\frac{e^{2}m^{\ast}a^{2}\Delta^{4}%
f^{3}\left(  \gamma_{0}\right)  }{h^{3}\left(  \mu^{\ast}B\right)  ^{2}%
kT}\sqrt[3]{\frac{m^{\ast}\Delta}{ah^{2}}}N.\label{GrindEQ__31_}%
\end{equation}
However, it is valid only for very strong magnetic fields. For instance,
for the parameters given above and ${kT/\Delta=0.03}$, the
longitudinal conductivity of the crystal in a magnetic field of 60~T amounts
to 0.675~$\Omega^{-1}\mathrm{m}^{-1}$, i.e. its resistance becomes about 1600
times as high as the resistance in the absence of a magnetic field. Hence, we
obtained that, in ultraquantum magnetic fields, the magnetoresistivity
obeys the law $\rho_{zz}\propto TB^{2}$. This law includes the
dependence on both the field and the temperature and can
be used for the experimental verification of whether the approximations made in this
paper and the corresponding model calculations were correct.

However, this asymptotic law has a substantial shortcoming from the formal
point of view. According to the law, the total longitudinal resistance of the
crystal tends to zero as $T\rightarrow0$ in strong magnetic fields. It is
difficult to explain such a behavior from physical reasons, if we are in the
region of the charge carrier scattering by charged impurities (although this law
has no physically unreasonable consequences at real temperatures). Therefore,
the question arises: Does another asymptotic law free of this shortcoming exist?

Before answering this question, we note that the same problem on the
longitudinal conductivity in layered crystals was considered in work \cite{5}.
However, it was done only for crystals with highly open Fermi surfaces, i.e.
for which ${\zeta_{0}/\Delta\gg1}$. Moreover, the consideration was carried
out in the approximation ${\zeta_{0}/(\mu^{\ast}B)\gg1}$. The result obtained
in work \cite{5} for the constant component of the conductivity at low
temperatures, when ${\zeta_{0}/(kT)\gg1}$, practically coincides with our one
presented in work \cite{11} for the case of a constant relaxation time. However,
the dependence of the oscillating part of the longitudinal conductivity on the
magnetic field is slightly different. If we abstract from the Dingle factor
which was also taken into account in work \cite{5}, this difference is mainly associated in the
case of weak magnetic fields with the fact that the
oscillating part of the conductivity was considered under conditions slightly
different from those analyzed in work \cite{11} and this paper, namely,
${\Delta/(\mu^{\ast}B)\gg1}${ and }$\omega_{c}\tau_{0}\ll1$. In this case, the
Landau subbands are smeared very much due to the interaction between charge
carriers and a random potential created by charged impurities. Under those
conditions, the oscillating part of the longitudinal conductivity cannot longer be
expressed in terms of the squared longitudinal velocity of charge carriers, as was
done in work \cite{11} and is done in this paper.

Moreover, the authors of work \cite{5} considered only such magnetic
fields, in which the FS remains open even in the case
${\Delta/(\mu^{\ast}B)\ll1}$ and $\omega_{c}\tau_{0}\gg1$. Hence,
the technique developed in that work demands modifications for the
case of closed surfaces with regard for a
\textquotedblleft squeezing\textquotedblright\ of the FS under the action of
a strong ultraquantum magnetic field.

In this work, we do not make such a modification in the case of a
quasiclassical magnetic field, because the majority of experiments
dealing with galvano-magnetic phenomena in layered crystals are
carried out under the conditions ${\Delta/(\mu^{\ast}B)\gg1}$ and
$\omega_{c}\tau_{0}\gg1$ \cite{15,16,17}, when the approach
presented in work \cite{11} and this one is applicable, although the
Dingle factor is taken into consideration at the treatment of
experimental data, because it allows the relaxation time of charge
carriers to be determined directly, at least on the extreme FS
cross-sections. However, let us make the specified modification in
the case of strong ultraquantum magnetic fields. In work \cite{5},
it was shown that, if charge carriers are scattered at the random
potential of charged impurities, the longitudinal
electroconductivity of the crystal can be considered as a diffusion
one in the case where ${\Delta/(\mu^{\ast}B)\ll1}${,
}${\omega_{c}\tau _{0}\gg1}$, ${\zeta_{0}/\Delta\gg1}$,
${\zeta_{0}/(\mu^{\ast}B)\gg1}$ as well. Here, $\omega_{c}$ is the
cyclotron frequency. Longitudinal conductivity can be determined at
$T=0$ by the formula
%32
\begin{equation}
\sigma_{zz}=\frac{8\pi^{3}e^{2}m^{\ast}a\tau_{0}\Delta^{2}}{h^{4}}\left(
1-\cos\frac{\pi\zeta_{0}}{\mu^{\ast}B}\right).\label{GrindEQ__32_}%
\end{equation}
This formula differs from that presented in work \cite{11} for the
constant component of the electroconductivity for the open FS in that the
constant relaxation time $\tau_{0}$ is substituted by the time depending on
the magnetic field, and this dependence is completely determined by the
expression in parentheses. However, the authors of work \cite{5} recognized
that formula (\ref{GrindEQ__32_}) is inapplicable, when, due to a high FS
openness, the Landau level intersects the Fermi one, because the longitudinal
electroconductivity vanishes at such magnetic fields, the situation being
unphysical. However, the condition of strong FS openness is not obligatory for
a formula of the type (\ref{GrindEQ__32_}) to be correct, because, as was indicated
in paper \cite{5}, formula (\ref{GrindEQ__32_}) was obtained by solving the
kinetic Boltzmann equation for \textit{every} Landau subband, totally or
partially filled. Therefore, for this formula to be valid, it does not matter
how many Landau subbands are filled. Hence, formula (\ref{GrindEQ__32_}) will
be generalized to the case of ultraquantum fields and a closed FS in two ways:
considering the single Landau subband with the number $n=0$ as 1) completely
or 2) partially filled. In both those cases, the Fermi energy is not assumed
constant, but it is dependent on the magnetic field according to formulas
(\ref{GrindEQ__19_}) and (\ref{GrindEQ__21_}).

Following the first way, the quantity $\Delta$ in formula (\ref{GrindEQ__32_})
is substituted by the quantity $\frac{\Delta^{3}f^{2}\left(  \gamma
_{0}\right)  }{16\left(  \mu^{\ast}B\right)  ^{2}}$, and the quantity
$\zeta_{0}$ by quantity (\ref{GrindEQ__21_}). Then, taking formulas
(\ref{GrindEQ__26_})--(\ref{GrindEQ__28_}) into account, we obtain the
following final expression for the longitudinal conductivity:
%33
\begin{equation}
\sigma_{zz}=\frac{\pi^{2}e^{2}m^{\ast}a^{2}\Delta^{5}f^{4}\left(  \gamma
_{0}\right)  }{64h^{3}\left(  \mu^{\ast}B\right)  ^{4}}\sqrt[3]{\frac
{12\pi^{2}m^{\ast}\Delta}{ah^{2}}}N. \label{GrindEQ__33_}%
\end{equation}
For the parameters given above and a magnetic field induction of 60~T,
the crystal conductivity amounts to $2.124\times10^{-5}~\Omega^{-1}$%
\textrm{m}$^{-1}$. Thus, if the single Landau subband is considered as
completely filled, the asymptotic law $\rho_{zz}\propto B^{4}$ is obtained.

Following the second way, let us use the general formula (\ref{GrindEQ__4_}), in
which the relaxation time is determined identically to how it was done in the
first way, and the integration over $x$, in accordance with
Eq.~(\ref{GrindEQ__19_}), is carried out from 0 to $\frac{f\left(  \gamma
_{0}\right)  \Delta}{2\mu^{\ast}B}$, with the integrand being restricted to
the square-law approximation with respect to $x$. Then, the longitudinal
conductivity of the crystal looks like
%34
\begin{equation}
\sigma_{zz}=\frac{\pi e^{2}m^{\ast}a^{2}\Delta^{4}f^{3}\left(  \gamma
_{0}\right)  }{3h^{3}\left(  \mu^{\ast}B\right)  ^{3}}\sqrt[3]{\frac{12\pi
^{2}m^{\ast}\Delta}{ah^{2}}}N.\label{GrindEQ__34_}%
\end{equation}
For the above-given parameters of the problem and a magnetic field induction of 60~T,
the crystal conductivity amounts to $5.018\times10^{-3}~\Omega^{-1}$%
\textrm{m}$^{-1}$. Thus, if the single Landau subband is considered as
partially filled, the asymptotic law $\rho_{zz}\propto B^{3}$ is obtained.
Hence, in a magnetic field of 60~T, the crystal electroconductivity becomes
almost eight, in the first case, and almost six, in the second case, orders of
magnitude as low as that in the absence of a magnetic field. It is clear that
the results obtained are to be experimentally verified. However, the
overwhelming majority of available experiments for galvano-magnetic phenomena
in crystals with a superlattice were carried out for highly open FSes (see,
e.g., work \cite{17} and the relevant references therein).

\section{Conclusions}

Hence, we proved that the layered-structure effects are essential not only for
open, but also for closed FSes. At a constant concentration of charge
carriers, those effects are essential even in the range where the quasiclassical
approximation is valid. In addition, there exists an optimal interval of
the magnetic field induction, where those effects are most brightly pronounced. In
the range where the quasiclassical approximation is valid, they manifest themselves
as an increase of the relative contribution of Shubnikov--de Haas oscillations
to the total conductivity and as some phase retardation of the oscillations.
In this case, the field dependence of the chemical potential almost does not
influence the oscillation manifestation character. In the optimal interval
of magnetic fields, the field dependence of the chemical potential considerably
affects the character of manifestations of the layered-structure effect. Concerning
the Kapitsa effect, the inverse situation takes place: the layered-structure
effects diminish the coefficient of proportionality between the
magnetoresistance and the magnetic field induction. Moreover, the longitudinal
Kapitsa effect, provided the charge carrier scattering at impurities, can be
explained only if the FS squeezing under the action of a magnetic field, i.e.
the field dependence of the chemical potential, is taken into account. Hence,
crystals with closed FSes and high degrees of the filling of the miniband should also be
regarded as layered ones. In addition, at rigorous calculations of the longitudinal
electroconductivity, the FS extension in the magnetic field direction and the
dependence of this extension on the magnetic field induction have to be taken
into consideration. In strong magnetic fields, the following asymptotic laws
can be obtained, depending on the intensity of the charge carrier scattering by
charged impurities and the way of how the filling of the single Landau subband
is simulated: $\rho_{zz}\propto TB^{2}$, if the conductivity is considered as
a drift one; $\rho_{zz}\propto B^{3}$, if the conductivity is considered as a
diffusion one and the single Landau subband is partially filled; and
$\rho_{zz}\propto B^{4}$, if the conductivity is considered as a diffusion one
and the single Landau subband is filled completely. The field dependence of
the chemical potential is also essential for the explanation of those dependences.

\rezume{%
ЧИ ВИРАЖЕНІ ЕФЕКТИ ШАРУВАТОСТІ \\ПРИ ЗАМКНЕНИХ ПОВЕРХНЯХ
ФЕРМІ?}{П.В. Горський} {У статті на прикладі поздовжньої
електропровідності у квантуючому магнітному полі, перпендикулярному
до шарів, показано, що ефекти шаруватості можуть бути виражені не
лише у кристалах із сильно відкритими поверхнями Фермі (ПФ), як це
традиційно вважається, але й у кристалах із замкненими ПФ.
Розрахунки проведено в наближенні сталого часу релаксації. У слабких
магнітних полях ефекти шаруватості виражаються у відставанні
осциляцій Шубнікова--де-Гааза (ШДГ) за фазою і у деякому збільшенні
їх відносного внеску. В області сильних магнітних полів існує
оптимальний діапазон, в якому ефекти шаруватості
 виявляються у різко немонотонній залежності поздовжньої
 електропровідності від магнітного поля. Крім того,
показано, що ефекти шаруватості ведуть до зниження коефіцієнта
пропорційності між магнітоопором і індукцією у поздовжньому ефекті
Капіци. Розглянуто також поздовжню електропровідність шаруватих
кристалів в ультраквантових магнітних полях і показано, що залежно
від того, як моделюється заповнення єдиної підзони Ландау і від
того, розглядається поздовжня провідність як дрейфова, чи як
дифузійна, можна отримати такі закони зміни магнітоопору з магнітним
полем: $\rho _{zz} \varpropto TB^{2}$, $\rho _{zz} \varpropto B^{3}
$ та $\rho _{zz} \varpropto B^{4} $.}

\end{document}